\documentclass[nocleveref, english]{lni}
\pdfoutput=1
\usepackage[english]{babel}
\usepackage[
  backend=biber,
  style=LNI,
  maxbibnames=4
]{biblatex}
\usepackage{graphicx}
\usepackage{mathtools}
\usepackage{csquotes}
\usepackage{orcidlink}

\usepackage{booktabs}
\usepackage{multirow}
\usepackage{array}
\usepackage{float}
\usepackage{svg}

\usepackage{array}
\newcolumntype{L}[1]{>{\raggedright\let\newline\\\arraybackslash\hspace{0pt}}m{#1}}
\newcolumntype{C}[1]{>{\centering\let\newline\\\arraybackslash\hspace{0pt}}m{#1}}
\newcolumntype{R}[1]{>{\raggedleft\let\newline\\\arraybackslash\hspace{0pt}}m{#1}}

\newcounter{challenge}[section]
\newenvironment{challenge}[1][]{\refstepcounter{challenge}\par\medskip
\noindent \textbf{Challenge~\thechallenge. #1} \rmfamily}{\hfill $\square$\\\\}

\newcounter{direction}[section]
\newenvironment{direction}[1][]{\refstepcounter{direction}\par\medskip
\noindent \textbf{Direction~\thedirection. #1} \rmfamily}{\hfill $\square$\\\\}
    
\usepackage{nicematrix}
\usepackage{tikz}
\usetikzlibrary{fit}
\usepackage{rotating}

\usepackage{enumerate}

\usepackage{subcaption}

\usepackage{tabularx}

\begin{document}
\title[Process Mining for Unstructured Data: Challenges and Research Directions]{Process Mining for Unstructured Data: Challenges and Research Directions}
\author[Agnes Koschmider \and Milda Aleknonytė-Resch \and Frederik Fonger \and Christian Imenkamp \and Arvid Lepsien \and Kaan Apaydin \and Dominik Janssen \and Dominic Langhammer \and Tobias Ziolkowski \and Yorck Zisgen]
{%
Agnes Koschmider\footnote{University of Bayreuth, Group Business Informatics and Process Analytics, Wittelsbacherring 10, 95444 Bayreuth, Germany \\ \email{{firstname.lastname}@uni-bayreuth.de}}\and%
Milda Aleknonytė-Resch\footnote{Kiel University, Group Process Analytics, Hermann-Rodewald-Str. 3, 24118 Kiel, Germany\\
\email{{mar, ffo, ale, kap}@informatik.uni-kiel.de}} \and%
Frederik Fonger\footnotemark[2] \and%
Christian Imenkamp\footnotemark[1] \and%
Arvid Lepsien\footnotemark[2] \and%
Kaan Apaydin\footnotemark[2] \and%
Maximilian Harms\footnotemark[1] \and%
Dominik Janssen\footnotemark[1] \and%
Dominic Langhammer\footnotemark[1] \and%
Tobias Ziolkowski\footnote{GEOMAR Helmholtz Centre for Ocean Research, Wischhofstr. 1-3, 24148 Kiel, Germany\\
\email{tziolkowski@geomar.de}} \and%
Yorck Zisgen\footnotemark[1]
}
\authorrunning{Koschmider et al.}
\startpage{1} 
\editor{} 
\booktitle{} 
\year{}
\maketitle

\begin{abstract}
 The application of process mining for unstructured data might significantly elevate novel insights into disciplines where unstructured data is a common data format. To efficiently analyze unstructured data by process mining and to convey confidence into the analysis result, requires bridging multiple challenges. The purpose of this paper is to discuss these challenges, present initial solutions and describe future research directions. We hope that this article lays the foundations for future collaboration on this topic. 
\end{abstract}
\begin{keywords}
Process Mining  \and Unstructured Data \and Challenges \and Directions
\end{keywords}

    \section{Introduction} 
\label{introduction}

The volume of data is continuously increasing and the ability and demand to efficiently analyze the data has become even more crucial. 
Machine learning and data mining are suitable techniques and tools to efficiently process and analyze the data. Complementary to both techniques is \textit{process mining}~\cite{vanderaalstProcessMiningData2016}. Process mining is a promising approach to find additional patterns (e.g., in terms of causal effects or bottlenecks) in data and in that way to give new insights into the data that could not be directly found with techniques like machine learning or data mining. The insights from processes are given by means of events that have been tracked by information systems. Then, this event data that is structured within a log (i.e., an event log), is used as input to any process mining algorithm. Process mining allows both an analysis based solely on event logs as well as a comparison between (manually generated or as-is) process models and an event log reflecting the to-be processes. By means of process mining, patterns can be uncovered in data with the objective to reveal comprehensive insights into the end-to-end processes aiming to answer questions like \enquote{When did what happen and in what order?}, \enquote{When will something happen?}, \enquote{Are there deviations from how it should have happened?} or \enquote{Will there be any unforeseen events?}. A plethora of process discovery algorithms exist, mostly focusing on structured data (e.g., \cite{weijtersProcessMiningHeuristics2006,guntherFuzzyMiningAdaptive2007,leemansDiscoveringBlockStructuredProcess2013}) and relying on three requirements for the event log: a case ID, an activity name and a timestamp. However, many application scenarios that are based on unstructured data would benefit from process mining.

In this paper, data is referred to as unstructured when it is not organized in a scheme that enables the retrieval of information as required by the desired application~\cite{boulton2006analysis}, i.e., it does not fulfill the requirements of an event log for process mining.
Note that this does not mean that unstructured data lacks any structure at all~\cite{feldman2007text}.
Rather, data can be unstructured for process mining while being structured for other applications -- e.g., unprocessed video data lacks case IDs and activity names, but is highly suitable and structured for the task of determining the color of a pixel at a specific timestamp.\\
Unstructured data is a common data format, for instance in disciplines like engineering or life and natural sciences.
These disciplines have a high demand to identify anomalies and causalities in the data and thus, to receive answers to the questions that process mining can respond to. However, unstructured data such as images, audio, video, documents, social media, or sensor data has not been curated in a computer-accessible form fit for process mining and thus to be directly used for these disciplines. This raises the need for processing methods that transform these unstructured data into a format that is structured for process mining.
For instance, video data can be transformed into data structured for process mining through the choice of an appropriate data collection, activity extraction, and event abstraction techniques \cite{lepsienAnalyticsPipelineProcess}.
The overarching challenge for process mining on unstructured data is that such data collections are on a lower level of abstraction as commonly used data for process mining, which is close to the business level.
Classical process mining assumes that event data is totally ordered, discrete, correct, and accurate, in an isomorphic relation with individual activity executions, and, finally, at a symbolic level.
Meeting these requirements is challenging for unstructured data.
Thus, a direct application of process mining is not possible or would not lead to useful results.
Instead, a number of intermediary steps are necessary before applying process mining on unstructured data.

The purpose of this paper is to discuss challenges and research directions related to process mining on unstructured data, similarly as conducted for other fields like modeling~\cite{michael_quo_2023}. Generally, on one hand, the challenges relate to the quality of the data, and on the other hand to the process of the data analysis (i.e., how is and could the data be processed?). 
Against this background, the paper is structured as follows. The next section summarizes a use case to which we will refer throughout the paper. Section~\ref{Pipeline} describes the analysis pipeline for process mining on unstructured data, which we commonly apply in our research. The challenges are presented in Section~\ref{challenges} and are followed by research directions, which are summarized in Section~\ref{directions}. The paper ends with a conclusion.
    \section{Illustration of Process Mining on Unstructured Data on a Use Case}
\label{sec:use_case}
\begin{figure}[ht]
    \centering
    \includegraphics[width=\linewidth]{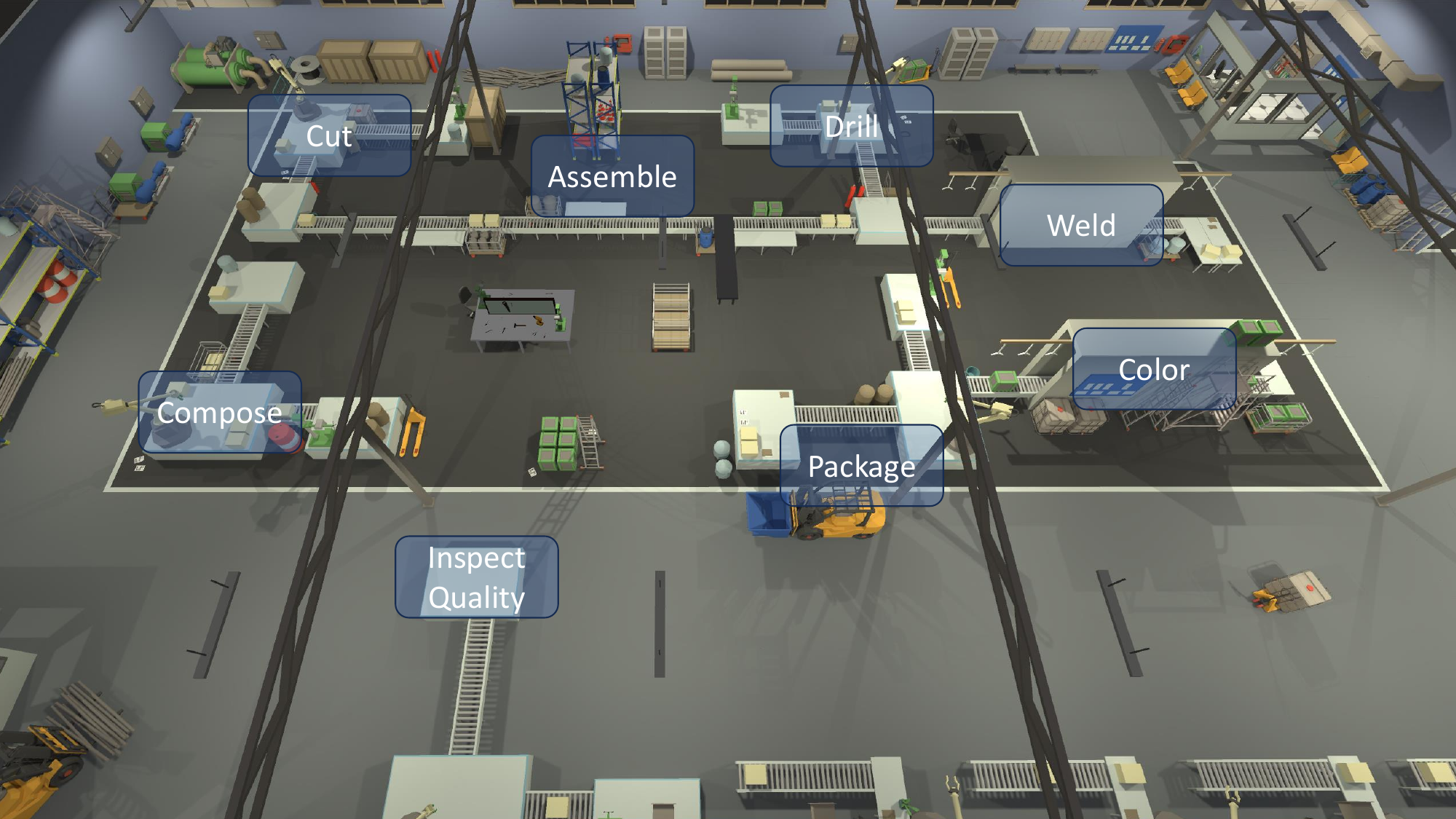}
    \caption{Smart factory use case with four assembly lines where each line is defined by tasks.}
    \label{fig:smart factory}
\end{figure}

This section introduces a smart factory use case, as seen in Fig, \ref{fig:smart factory}. We will refer to this use case throughout the paper to illustrate the challenge and future research directions for process mining on unstructured data. The use case refers to a production process with four assembly lines. One assembly line puts raw components together into a product. Subsequently, the next line divides them into a  drilling and welding step, accordingly. Then, the product is colored and packaged. Table \ref{tbl:example_eventlog} shows an excerpt of an exemplary event log with products, timestamps, and activities from such a production. Several sensors and video cameras are installed in the smart factory for automatic control. For instance, temperature, humidity, and carbon dioxide are tracked, while cameras monitor the tasks of the assembly line. Finally, the quality of the product is inspected. 

Using unstructured data for process mining for this use case requires efficient pre-processing techniques, which we will summarize in the next section. In the following, we distinguish between unstructured data, which has been captured regularly (like time series data) or irregularly (e.g., event-based). Time series data might be captured either as a continuous, irregular, or periodic stream within specific time frames. Sensors of a smart factory generate time series data through temperature or humidity sensors. The data has a low level of abstraction as raw sensor readings are stored in the data collection, as exemplary shown in Table~\ref{tbl:example_time_series_data_1}. This is different from business data which is typically used for process mining, which has activities at a high level of abstraction. Event-based data is captured when triggered by events, e.g., an alarm that is activated in response to an event (e.g., exceeding a temperature value). 

    \begin{table}[htp]
                        \centering
                        \renewcommand{\arraystretch}{1.2}
                        \renewcommand{\tabcolsep}{0.2cm}
                        \begin{tabular}{cccc}
                            \toprule
                             \textbf{Product ID} & \textbf{Timestamp} & \textbf{Activity} & \textbf{\dots} \\
                            \midrule 
            
                            1234 & 2021-01-01, 10:00 & Compose \\
                            1567 & 2021-01-01, 10:00 & Color  \\
                            1567 & 2021-01-01, 10:05 & Inspect Quality    \\
                            1567 & 2021-01-01, 10:10 & Package \\
                            1234 & 2021-01-01, 10:10 & Color \\

                            \dots & \dots & \dots \\
                            \bottomrule
                        \end{tabular}
                        \vspace{0.2cm}
                        \caption{Exemplary event log for the smart factor production use case (Fig.~\ref{fig:smart factory}).}
                        \label{tbl:example_eventlog}
    \end{table} 

        \begin{table}[t]
                \centering
            	\renewcommand{\arraystretch}{1.2}
            	\renewcommand{\tabcolsep}{0.2cm}
            	\begin{tabular}{ccc}
            	    \toprule
            		\textbf{Timestamp} & \textbf{Sensor Type \& ID} & \textbf{Sensor Value} \\
            		\midrule 
            		\dots & \dots & \dots \\
                	2021-01-01 12:59:57 & \texttt{MotionSensor22} & ON \\
                	2021-01-01 13:06:21 & \texttt{MotionSensor22} & OFF \\
                	2021-01-01 13:06:22 & \texttt{MotionSensor14} & ON \\
                	2021-01-01 13:06:23 & \texttt{TemperatureSensor04} & $22$ \\
                	2021-01-01 13:06:24 & \texttt{MotionSensor23} & OFF \\
                    2021-01-01 13:06:23 & \texttt{TemperatureSensor07} & $65$ \\
            	    \dots & \dots & \dots \\
            	    \bottomrule
            	\end{tabular}
            	\vspace{0.2cm}
            	\caption{Example of stationary sensor event data}
            	\label{tbl:example_time_series_data_1}

        \end{table}

Process mining can give valuable insights into the smart factory use case in terms of discovering reasons for, e.g. late packaging and deliveries, identifying bottlenecks, or improving predictions. In addition to the smart factory use case, process mining on unstructured data might provide also benefits in domains such as healthcare, logistics, finance, and education.
    \section{The Process Analytics Pipeline}
\label{Pipeline}

\begin{figure}[htb] 
    \centering
    \includegraphics[width=\linewidth]{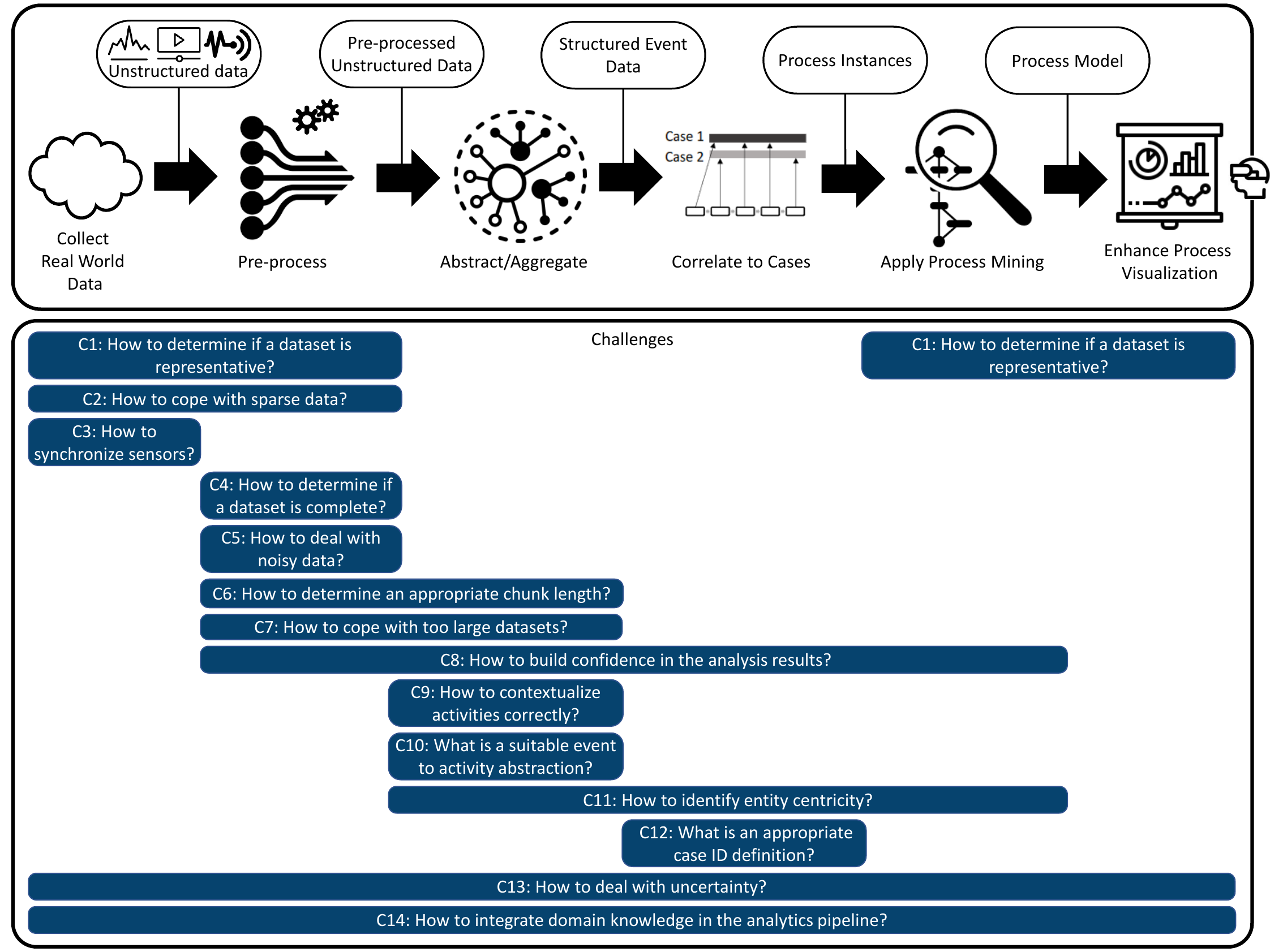}
    \caption{The way from unstructured data to process visualization: the steps to get there.}
    \label{fig:processanalytics}
\end{figure}

To efficiently analyze data, requires a systematic approach. Fig.~\ref{fig:processanalytics} shows such an approach in terms of a process analytics pipeline. The pipeline consists of five subsequent steps, which generally apply to all data formats, structured and unstructured. However, applying this pipeline requires a prior evaluation if new insights can be gained for the specific use case, particularly when unstructured data is involved. Additionally, each step in the pipeline needs to be adjusted to the case and its specific characteristics (e.g., when the data is distributed and distributed analysis is an efficient analysis solution). The subsequent paragraphs summarize each step of the pipeline, focusing on the handling of unstructured data in the smart factory use case.

\paragraph{Pre-processing}
Data pre-processing is a time-consuming and error-prone task due to extensive tasks involved like data integration, data enhancement and enrichment, data transformation, data reduction, data discretization and data cleaning~\cite{10.1145/3613247}. Data pre-processing is necessary before the data from the raw format can be used (i.e., time series data must be transformed into an appropriate data format), noise and outliers (e.g., erroneous or missing values) must be removed, and representative data has to be extracted. In the smart factory, video data could be pre-processed in order to delete images of low resolution, adjust frame rates, and label data with bounding boxes and activity names~\cite{lepsienAnalyticsPipelineProcess}.
The processed data must be divided into subsequences representing time periods (like seasons, or weeks). Clustering is one appropriate technique to find similarities between subsequences. As a result, one subsequence can be assigned to one cluster, while one cluster is mapped onto a process activity. We call the step from a cluster to an activity an event/activity abstraction, which is summarized next. The data shown in Tab. 2 might be used as input for the pre-processing step.

\paragraph{Abstraction} Next, raw data has to be enriched with semantics. Assume that the raw data shown in Tab. 2 has been used as input for the abstraction phase, then aggregation techniques are needed to leverage the semantics of raw data to a higher level of abstraction. For this, several techniques exist to abstract high-level events (e.g., Table \ref{tbl:example_eventlog}) from pre-processed raw data (e.g., Table \ref{tbl:example_time_series_data_1}). A comprehensive survey on event abstraction for structured data was conducted by van Zelst et al.~\cite{van_zelst_event_2021}. High-level events for process mining can be abstracted through thresholds between data entries (e.g., between temperature measurements for machines in the smart factory to define activities such as \enquote{temperature increases}, therefore classification of time series data can be use for the abstraction of activities), by applying knowledge gathered from domain experts \cite{janssenProcessMiningSensor,lepsienAnalyticsPipelineProcess} or by using machine learning to uncover more complex patterns \cite{bishop2006pattern,murty2015introduction}.

\paragraph{Case Correlation}
In this step, the abstracted (high-level) events have to be correlated to process instances by assigning case IDs.
Several properties have been used to relate one record to an instance of a process and often techniques must be designed allowing to involve domain knowledge or make assumptions about the start and end of process instances~\cite{laueProzessmanagementUndProcessMining2020,DBLP:conf/zeus/LepsienBMK22,janssenProcessMiningSensor,janssenProcessModelDiscovery2019}. In the smart factory, activities could be assigned to cases by setting the case ID equal to the product ID of the product (see Fig. 1) that is being produced such that one instance of the process begins with the collection of raw material and ends with the packaging of the final product. Different process discovery algorithms exist to extract process models from event data, each with advantages and disadvantages~\cite{laueProzessmanagementUndProcessMining2020}, some of the most popular being the Heuristic Miner \cite{weijtersProcessMiningHeuristics2006}, Fuzzy Miner \cite{guntherFuzzyMiningAdaptive2007} and Inductive Miner \cite{leemansDiscoveringBlockStructuredProcess2013}. To analyze the event data of the smart factory, discovery algorithms could be applied to mine models of the most frequent production steps or get a holistic overview of all process variations.

\paragraph{Visualization} The visualization of event data, as well as the visualization of process graphs can be approached with different diagrams and models for analysis. The visualization of the process graphs with e.g., Petri nets allow the analysis of the control flow, i.e. the order of activities, choices, concurrency, and loops within the process. Further, process graphs can be augmented in order to analyze how often activities follow one another or to analyze the duration of activities in terms of time differences between activity executions. Moreover, process mining tools are able to simulate the discovered process models, which allow additional insights \cite{vanderaalstProcessMining3602022}. Futhermore, initial studies show that the adaption of 3D visualizations, e.g., virtual or augmented reality, can have benefits due to the visualization space and depth \cite{wetzel2022}. Besides process graphs, additional visualization techniques exist, such as dotted charts \cite{songSupportingProcessMining2007}, sequence diagrams \cite{leemansHierarchicalPerformanceAnalysis2018}, heatmaps \cite{lepsienAnalyticsPipelineProcess} 
and decision trees \cite{deleoniGeneralFrameworkCorrelating2014}.
In the smart factory use case dotted charts might be used to visualize when production steps are executed.
    \section{Challenges}
\label{challenges}

This section presents challenges that need to be addressed to advance the area of process mining on unstructured data. These challenges stem from various use cases we have encountered e.g. \cite{DBLP:conf/BPMDS/Ziolkowski2022, janssenProcessMiningSensor, lepsienAnalyticsPipelineProcess, melfsen_describing_2023}. The challenges lay the foundation to formulate research directions that we will describe in Section~\ref{directions}. To align the challenges with the process analytics pipeline, we cross-reference the challenges with steps of the pipeline~(Fig.~\ref{fig:processanalytics}).

\begin{challenge}
\label{challenge:representativeness}
        How to determine if a dataset is representative? 
\end{challenge}
The representative selection of a dataset impacts the goal of the analysis. For example, if the goal of the analysis is anomaly detection, then the anomalous behavior results from the definition of anomalous vs. behavior that is within the range. The analysis goals should be fixed before gathering data and thus, the data collection is planned in such a way as to collect representative data. In practice process mining is rarely the primary goal for gathering data, but rather, already existing data needs to be pre-processed to fit the requirements for process mining. Thus, it is imperative to assess the suitability of the dataset for the intended analysis. An exploratory data analysis might help to get initial insights whether the gathered data aligns with the analysis goals and is not a \enquote{garbage-in-garbage-out} approach \cite{Wynn}. Assume that the analysis goal is to find anomalies in the first assembly line, then the observation period would have to be long enough to collect data affecting all steps of the first assembly line. If the data set is considered to be non-representative, it will not necessarily impact the data abstraction and case correlation pipeline steps, but will affect the conclusions made from the analysis (i.e., lead to misleading conclusions).

\begin{challenge}
\label{challenge:sparseness}
    How to cope with sparse data?
\end{challenge}
Process data has to be captured for processes that are frequently repeated so that variations are detected and can be analyzed \cite{grisoldAdoptionUseManagement2020a}. For example, in a smart factory, sensors that are attached to machines could be installed for a limited time period, which could lead to too sparse data for process mining. To overcome this challenge, related data could be synthetically generated \cite{zisgen_generating_2022}. Depending on the type of data, this could be done by using algorithms that generate additional data, or by applying noise to a given data set \cite{goodfellow2016deep} during the pre-processing step (Fig. \ref{fig:processanalytics}). Alternatively, more real-world data may need to be collected.

\begin{challenge}
\label{challenge:synchronization}
    How to synchronize sensors?
\end{challenge}
Time synchronization is an issue in case multiple sensors are in use and interact with each other~\cite{van_eck_enabling_2016}. If sensors are not synchronized in the data collection step (Figure \ref{fig:processanalytics}), it might result in temporal discrepancies in the recorded data as well as incorrect representations of the behavior of events and finally, lead to erroneous interpretations of the process models derived from the event log. For example, multiple motion detection sensors monitoring different areas of the smart factory must be calibrated for the same timestamp in order to correctly identify moving entities from one area to another.

\begin{challenge}
\label{challenge:completeness}
    How to determine if a data set is complete?
\end{challenge}
A data set may not be complete due to missing data. Reasons for missing values are unreliable sensors that do not always work as intended, sensors that were not used properly, or if data recording was not possible. For instance, assume that wearable sensors are used, but have not been turned off while wearing and thus continuously generate data. Also, it is challenging to identify rare events in the data since it is not clear whether the event is missing or rare. To counteract this, counting event frequencies or applying rare event detection techniques might be a solution  \cite{Harrison}. Referring to the smart factory example, let's assume that location detection is an analysis purpose. Then external data about power failures could be included beside the data from the movement detection sensors to determine data set completeness. Incomplete data can lead to erroneous data logs and thus impact not only the abstraction/aggregation pipeline step, but all other following steps.

\begin{challenge}
\label{challenge:noise}
    How to deal with noisy data? 
\end{challenge}
Logged data can have different types of imperfections \cite{suriadiEventLogImperfection2017}, especially incorrect data \cite{bose}. Therefore, the quality of the data should be improved during the pre-processing step (see Fig. \ref{fig:processanalytics}) using data cleaning techniques \cite{janssenProcessMiningSensor}. Outliers can lead to the creation of models that contain a significant number of infrequent execution paths or that do not accurately reflect the behavior \cite{confortiFilteringOutInfrequent2017}. For example, imagine tracking the location of an object in the smart factory. Even though the object is still at an intermediate production step, i.e., further steps within the production line are mandatory, the captured location records could indicate a transport to the warehouse. This occurrence might be tracked back to software or sensor failure. Filtering such cases becomes necessary. Otherwise, the final process model might contain paths that are not possible in the real world. Furthermore, there are also cases when filtering techniques are not applicable. For example, when a video camera in the smart factory captures data that contains scenes where the images are blurred due to unexpected light, movement, or a malfunctioning focus. In such a case it might be more efficient to proceed with the recording of another data set instead of handling the erroneous data. Eventually, the step of the process analytics pipeline \enquote{pre-processing} is often a labor-intensive task but has a significant impact on the process mining result.

\begin{challenge}
\label{challenge:chunk_size}
    How to determine an appropriate chunk length?
\end{challenge}
To determine the chunk length of unstructured data is crucial for several reasons. Firstly, the chunk length determines the granularity at which the data is analyzed. Depending on the goal of the analysis, a smaller chunk length allows detailed insights into the analysis goal. Contrary, a larger chunk length allows a more general view.           
Also, the chunk length has an impact on clustering, classification, and pattern recognition \cite{janssenProcessModelDiscovery2019}. A sliding window technique in the pre-processing step (Fig. \ref{fig:processanalytics}) can be used to divide the data into overlapping chunks of varying sizes, allowing for analysis of different resolutions and improved identification of patterns and trends. Statistical techniques such as power analysis and sample size calculation can also be employed to determine an appropriate chunk length. Alternatively, machine learning algorithms, including clustering and dimensionality reduction techniques such as Principal Component Analysis, can be applied to extract meaningful information from high-dimensional time series data and improve interpretability for further analysis. In a smart factory, sensors that are installed in machines and in the facility produce big data. This amount of data needs to be segmented into a chunk size appropriate for the research question. Finally, processing large chunks of unstructured data can be computationally intensive. By determining an appropriate chunk length, data processing times can be reduced. Thus, the chunk length selection is a trade-off between computational complexity and information accuracy related to the analysis goal. 

\begin{challenge}
\label{challenge:volume}
    How to cope with too large datasets?
\end{challenge}
If the data volume is too high, computational complexity is an issue affecting all steps of the process analytics pipeline (see Challenges \ref{challenge:chunk_size} and \ref{challenge:events}).  
Re-scaling or sampling allows reducing the quantity of the data~\cite{sayood2017introduction,lepsienAnalyticsPipelineProcess}, which in our case would be part of the pre-processing step. However, this could impact the amount and accuracy of activities abstracted from the events. Thus, when reducing the amount of data, not only the processing time but also the meaningfulness of abstraction needs to be considered \cite{lepsienAnalyticsPipelineProcess}. As an example, video data gathered from cameras in the smart factory may require rescaling as a specific measure to manage data volume. The resolution of the videos might need to be scaled down to reduce file size until processing time becomes adequate while activities, e.g., distinct manual assembly steps, remain detectable \cite{egmont2002image,ma2019image,hojjat2023progdtd}. Thus, given such scenarios, the applicability of process mining techniques becomes limited. 

\begin{challenge}
\label{challenge:confidence}
    How to build confidence in the analysis results?
\end{challenge}
While process mining on unstructured data can produce valuable insights, its potential to realize this value is inherently limited by the confidence that decision-makers place in it \cite{mehdiyev_explainable_2021}.
Providing confidence in the process mining results, which rely on machine learning techniques (e.g., for object and activity recognition from video recordings in the smart factory) is challenging due to the black-box nature of these models \cite{mehdiyev_explainable_2021}. These concerns are further amplified when multiple pipeline steps are required to prepare the data, raising the need for methods to communicate the results in a way that builds trust in the analysis and confidence in the decisions based on the results \cite{DBLP:journals/insk/KoschmiderOH22}.

\vspace{1cm}

\begin{challenge}
\label{challenge:contextualization}
    How to contextualize activities correctly?
\end{challenge}
The methods available for pre-processing unstructured data for process mining usually do not integrate the context of the analyzed process. This can be problematic, as two activities that are distinct in a process may be difficult to distinguish at the raw data level \cite{rebmann_enabling_2019}.
For instance, in the smart factory example, wearable sensors would produce very similar data for lifting a workpiece from one workstation and putting it down at the next station. Because the pre-processing steps miss contextual information, they are unable to efficiently distinguish between these activities during the abstraction/aggregation step. Thus, in the pipeline steps after pre-processing, the data needs to be explicitly contextualized into the realm of the analyzed process.

\begin{challenge}
\label{challenge:events}
    What is a suitable event to activity abstraction or aggregation? 
\end{challenge}
The challenge of abstraction or aggregation particularly when processing unstructured data is that semantics must be put into input data that is generally not understandable at all (i.e., the input data is at a lower level of abstraction than business data). The abstraction of activities to events has been extensively studied~\cite{van_zelst_event_2021}. A too fine-grained abstraction leads to the overfitting of the discovered process model, while a too coarse-grained abstraction results in underfitting.
To illustrate this, consider the activity of cutting wood. Then the question arises if the start of the activity is initiated when the wood enters the area of the machine, when the machine or the lid starts running or is closed, or when the cutting process starts. These questions need to be answered before starting the analysis and the same requirement applies to all data anylsis and data sets. The abstraction of activities from events (in terms of granularity) depends on the analysis purpose~\cite{van_zelst_event_2021, bi_event_2018}.

\begin{challenge}
\label{challenge:entity_identification}
    How to identify entity centricity? 
\end{challenge}
Most stationary sensors operating in the event-based capture mode and video cameras share the challenge of identifying the observed entities, especially if a lack of identification tags is present.
Regarding video data, in the case of observing humans, entities could be re-identified using facial recognition~\cite{kratsch_shedding_2022} during pre-processing, abstraction, and aggregation as well as case correlation. However, there are monetary and sensor management trade-offs when implementing these strategies. Similarly with video data, if the entities exit the field of view, it is difficult to determine which entity has re-entered the observation field first. For a different example, a smoke detector placed in the middle of the smart factory cannot identify the machine or entity triggering it. By increasing the number of sensors or video cameras present, it may be easier to identify the entities, e.g., in the case of a smoke detector, a number of smoke detectors could be placed above each factory machine, and thus the smoke detector right above the machine emitting smoke will be triggered first. 

\begin{challenge}
\label{challenge:cases}
    What is an appropriate case ID definition? 
\end{challenge}
A further challenge when preparing the data for process mining is the definition of the case ID, which is necessary for process mining algorithms to extract process models from event logs. The case notion might be ambiguous and thus selecting and assigning an appropriate case ID during correlation is challenging since no distinction of cases exists, e.g., in time series data from natural and life science \cite{melfsen_describing_2023}. Also, applications exist where the systems are not able to log the case IDs \cite{bayomie-event-case-2023}. Within the smart factory, sensors attached to some production steps may generate a continuous data stream (e.g. torque, temperature, humidity) which does not include identifiers to correlate their output with steps in the production cycle of this specific production step. After defining activities and events through an abstraction process (see Challenge\ref{challenge:events}), case IDs have to be defined artificially.

\begin{challenge}
\label{challenge:uncertainty}
    How to deal with uncertainty? 
\end{challenge}
Generally, the collection and processing of unstructured data are subject to errors, which introduces uncertainty to the analysis. Because this uncertainty is inherent in every analysis using process mining on unstructured data, uncertainty-aware process mining techniques, i.e., techniques that explicitly handle the uncertainty attached to an event log, and quantify or visualize it, are required \cite{buchs_proved_2021,pegoraro_probabilistic_2022}. For instance, inadequate data collections (e.g., due to too low video frame rates, incorrect sensor placements), and the probabilistic mappings applied during the abstract/aggregate and correlate to case steps, lead to event logs not being fully significant \cite{koschmider2019contextualization}. 

\begin{challenge}
\label{challenge:domain_knowledge}
    How to integrate domain knowledge in the analytics pipeline?
\end{challenge}
While many steps of the analytics pipeline can be fully automated (e.g., collecting real-world data, pre-processing), the main challenge is understanding the practical application contexts where domain knowledge is crucial. Domain knowledge can be provided in many different forms, e.g., textual descriptions of a process or relationships between activities, formal constraints, or normative process models. For example, domain knowledge can be used to facilitate event abstraction \cite{baier_matching_2018}, in case correlation \cite{bayomie_correlating_2016} or to repair missing events in traces \cite{bogdanov_sktr_2023}. The focus here is recognizing and defining the areas where domain knowledge integration can enhance analysis quality.

    \pagebreak
    \section{Future Directions}
\label{directions}
This section summarizes research directions that should be tackled in the future in order to significantly advance the field of process mining on unstructured data. 

\begin{direction} 
\label{direction:domain_knowledge}
    Integration of domain knowledge.
\end{direction}
\textit{Related challenges}:  \ref{challenge:representativeness}, \ref{challenge:contextualization}, \ref{challenge:entity_identification}, \ref{challenge:cases}, \ref{challenge:domain_knowledge}

Domain knowledge is pivotal for understanding the context of processes and events. Integrating this knowledge into the analytics pipeline is crucial (see Fig. 1). A core aspect of this direction is to set the focus on the development of \textit{generic} methods and models for integrating domain knowledge into the analytics pipeline. Techniques such as natural language processing and sentiment analysis facilitate the extraction of vital information from textual data. Analyzing language in documents and other unstructured sources give additional insights and uncover patterns not evident in structured data alone.

A key aspect of this direction is designing hybrid models that combine data science methods with human input, which can integrate this domain knowledge into specific use cases. Here, humans give domain-specific insights to enhance the analysis's accuracy. For instance, factory workers can provide additional information about the production process and the relationships between different activities that can be used to improve the accuracy of the analysis, while domain experts can shed light on particular challenges related to smart factories.

Future studies should emphasize hybrid models for seamless domain knowledge integration. This entails designing techniques to capture and weave into domain-specific insights, implementing algorithms to assimilate this knowledge efficiently, and establishing validation protocols. Moreover, integrating feedback loops will be the key to refining these models sustainably.

\begin{direction} 
\label{direction:data_fusion}
    Exploration of data fusion techniques for comprehensive analysis.
\end{direction}
\textit{Related challenges}: \ref{challenge:sparseness}, \ref{challenge:synchronization}, \ref{challenge:completeness},  \ref{challenge:volume}, \ref{challenge:events}

Data fusion combines structured and unstructured data sources and gives additional insights into the analysis. This involves integrating data from distributed sources to create a more complete and accurate picture of the underlying processes.
For instance, in the smart factory use case, data from sensors and video cameras are collected to monitor the tasks of the assembly line.
By fusing this data, and other structured data sources such as production logs and quality control reports, it is possible to identify patterns that may not be apparent from solely one data source alone. 
This requires the development of new techniques for data pre-processing, feature extraction, and data integration for heterogeneous data. 
To enable applications such as predictive monitoring and support during execution, these data fusion techniques need to be developed with a focus on scalability to enable integration and analysis of the data in real-time.

\begin{direction} 
\label{direction:advanced_visualization}
    Use of advanced visualization techniques for unstructured data. 
\end{direction}
\textit{Related challenges}:  \ref{challenge:noise}, \ref{challenge:chunk_size}, \ref{challenge:confidence}

New visualization techniques are required to efficiently convey the results of process mining on unstructured data to stakeholders. Traditional process mining techniques often rely on visualizations in terms of process models. These visualizations may not always be suitable for unstructured data sources, which require different kinds of visualization views. It is imperative to strike a balance between complexity and clarity: it must be complex enough to convey the insights and at the same time simple enough to be understandable. For instance, in the smart factory use case, the data collected from sensors and video cameras can be visualized using interactive dashboards, and 3D visualizations, among others. 

Furthermore, the scalability of advanced visualization techniques will emerge as a paramount requirement. As the volume and variety of data sources increase, it becomes more challenging to visualize the data in real-time. Therefore, future research should focus on developing scalable visualization techniques that can handle large volumes of data from different sources.

\begin{direction} 
\label{direction:explainability}
    Use of machine learning to make process mining on unstructured data more explainable.
\end{direction}
\textit{Related challenges}: \ref{challenge:noise}, \ref{challenge:volume}, \ref{challenge:confidence}, \ref{challenge:entity_identification}

As process mining on unstructured data becomes more prevalent, it is important to ensure that the results are explainable and trustworthy. This is particularly important when using unstructured data in critical fields, such as medicine or law. Otherwise, reliable decisions could be compromised, which could lead to incorrect medical treatments or unfair legal outcomes. Future research should tackle explainability by developing machine learning models that enhance data accuracy (e.g., via feedback mechanisms) while incorporating explainability. 

\begin{direction} 
\label{direction:ethical_legal_implications}
    Research and develop frameworks for ethical and legal implications.
\end{direction}
\textit{Related challenges}: \ref{challenge:confidence}, \ref{challenge:uncertainty}

Collecting and processing unstructured data such as camera images or sensor readings raises issues related to data protection and privacy, and possibly bias and discrimination if the data is not collected properly, undermining confidence in the results. Furthermore, when decisions from the analysis are drawn, people in less represented cases could be discriminated against \cite{van_der_aalst_responsible_2022}.

Also, process mining currently lacks clear guidelines and regulations for using unstructured data. Therefore, future research should focus on developing comprehensive frameworks for ethical, legal, and transparent data governance. This involves examining the ethical and legal issues involved in using unstructured data for process mining and developing guidelines and regulations to ensure that the data is collected, stored, and analyzed in a responsible and accountable manner. To guarantee privacy, techniques need to be developed that can anonymize the data while preserving its utility. This is an area that requires further research~\cite{DBLP:journals/tmis/ElkoumyFSKMVRW22}.

A collaboration between researchers, practitioners, and policymakers is needed to develop these frameworks. This requires a deep understanding of the ethical and legal issues involved in process mining and the ability to transfer this knowledge into practical guidelines and regulations.
    \section{Conclusion}

This paper presented challenges and research directions related to process mining on unstructured data. The quality of the data and the process of data analysis are two major challenges that need to be addressed. To overcome these challenges, an analysis pipeline consisting of five subsequent steps has been presented. The challenges presented arise from the large number of our practical experiences with process mining on unstructured data.

Process mining on unstructured data is a challenging but promising area of research. By addressing the challenges and exploring new research directions, the potential of unstructured data can be unlocked, and new insights into processes can be gained that traditional techniques might not fully respond.

\section*{Acknowledgments}
This project has received funding from the State of Schleswig-Holstein under the Datencampus project grant no. 220 21 016, the German Research Foundation (DFG) SPP 2422 and FOR 5495, the Federal Ministry for Digital and Transport under the CAPTN-Förde 5G project grant no. 45FGU139 H, the Federal Ministry for Economic Affairs and Climate Action under the MARISPACE-X project grant no. 68GX21002E and the German Federal Ministry of Education and Research (BMBF) for the ABBA project grant no. 16DHBKI002, 16DHBKI003, 16DHBKI004, 16DHBKI005.


\printbibliography
\end{document}